\documentclass[aps,prd,twocolumn,showpacs,preprintnumbers,amsmath,amssymb]{revtex4-1}
\usepackage{graphicx}
\usepackage[usenames]{color}
\begin{document}

\preprint{FERMILAB-PUB-13-394-T}

\title{Controlling sign problems in spin models using tensor renormalization}
\author{Alan Denbleyker$^1$}
\author{Yuzhi Liu$^{1, 2, 4}$}
\author{Y. Meurice$^1$}
\author{M. P. Qin$^3$}
\author{T. Xiang$^3$}
\author{Z. Y. Xie$^3$}
\author{J. F. Yu$^3$}
\author{Haiyuan Zou$^1$}
\affiliation{$^1$ Department of Physics and Astronomy, The University of Iowa, Iowa City, Iowa 52242, USA }
\affiliation{$^2$ Theoretical Physics Department, Fermi National Accelerator Laboratory, Batavia, Illinois 60510, USA}
\affiliation{$^3$ Institute of Physics, Chinese Academy of Sciences, P.O. Box 603, Beijing 100190, China}
\affiliation{$^4$ Department of Physics, University of Colorado, Boulder, CO 80309, USA }

\date{\today}
\begin{abstract}
We consider the sign problem for classical spin models at complex $\beta =1/g_0^2$  on $L\times L$ lattices. We show that the tensor renormalization group method allows reliable calculations for larger Im$\beta$ than the reweighting Monte Carlo method. For the Ising model with complex $\beta$ we compare our results with the exact Onsager-Kaufman solution at finite volume. The Fisher zeros can be determined precisely with the TRG method. We check the convergence of the TRG method for the $O(2)$ model on $L\times L$ lattices when the number of states $D_s$ increases. We show that the finite size scaling of the calculated Fisher zeros agrees very well with the Kosterlitz-Thouless transition assumption and predict the locations for larger volume. The location of these zeros agree with Monte Carlo reweighting calculation for small volume. The application of the method for the $O(2)$ model with a chemical potential is briefly discussed. 
\end{abstract}

\maketitle

\section{Introduction}

Sign problems appear generically in models for fermions with a chemical potential. Despite the importance of calculations at finite density in many situations, this has remained a very challenging problem for Monte Carlo (MC) practitioners in particle physics and condensed matter physics \cite{regensburg}. Sign problems also occur in models with complex couplings or temperature. At finite volume, lattice models with compact field variables usually have a partition function which is analytical in the entire complex coupling plane. Studying the analytical continuation of these models  can be used to understand the convergence of various types of expansions \cite{falcioni81} and to distinguish different types of phase transitions \cite{janke00,janke04}. Complex renormalization group flows can 
also be defined and the zeros of the partition function in the complex coupling plane, called Fisher zeros, determine their global structure \cite{flow10prl,Onflow11,Hmodelflow11}. 

Calculations with a complex coupling can be performed by reweighting  MC configurations generated without sign problem using a real coupling. A simple example is the calculation of the partition function of a spin or gauge model with a complex inverse coupling $\beta={\rm Re}  \beta+i{\rm Im}\beta$. It can be expressed as \cite{falcioni82}:
\begin{equation}
\label{eq:normpart}
Z({\rm Re} \beta+i{\rm Im}\beta)/Z({\rm Re} \beta)=<{\rm e}^{-i{\rm Im}\beta E}>_{{\rm Re} \beta} \ .
\end{equation}
However, the fluctuations of this quantity become of the same size as the average when  ${\rm Im} \beta$ is too large. For a Gaussian distribution of energies (actions), this occurs when ${\rm Im}\beta$ of the order of $V^{-1/2}$ \cite{Alves1992nupb}. 

In this article, we argue that the Tensor Renormalization Group (TRG) method for classical models \cite{Nishino95,Levin2007,GuLevinWen08,XiePRL09,ZhaoPRB10,XiePRB12,YMPRB13,efrati13,Exactblocking13prd,Yuxy13} provides a solution to the sign problem preventing direct MC simulations at complex $\beta$. This method can be applied to most models \cite{YMPRB13,Exactblocking13prd} studied by lattice gauge theorists.
In the following, we use the TRG method based on the higher-order singular value decomposition (HOTRG) \cite{XiePRB12} to calculate the complex partition functions of the two-dimensional (2D) classical Ising and $O(2)$ (also called classical XY) models with complex $\beta$. The reason why the HOTRG method is relatively insensitive to complex values of $\beta$ is explained  in the next section (Sec. \ref{sec:method}).    

The rest of this paper is organized as follow. In Sec. \ref{sec:ising}, we consider the exactly solvable case of the 2D Ising model on a square lattice. A highly accurate numerical solution from HOTRG calculations at real temperature has already be obtained in Ref \cite{XiePRB12}. We extend the HOTRG calculation to the partition function at complex $\beta$ for finite volumes and determine the zeros of the complex partition functions. By increasing the number of states $D_s$ used in the HOTRG procedure, the agreement with the exact Onsager-Kaufman solution \cite{Kaufman1949is} improves and the error is much smaller than the MC reweighting method. However, for a chosen accuracy, $D_s$ must be increased as one approaches a Fisher zero. Next, in Sec. \ref{sec:xy}, we apply the same method to search for the zeros of partition functions for the 2D $O(2)$ model at finite volume and analyze the lowest zeros for different volumes  by using finite size scaling. This yields the analytic behavior of the zeros at the large volume limit. Finally, results are summarized and ongoing practical applications with a chemical potential are discussed in Sec. \ref{sec:conl}.   

\section{HOTRG}
\label{sec:method}
The partition function of a spin or gauge model with local interactions can be represented as the trace of a product of local tensors. For the 2D classical Ising and $O(2)$ models on a square lattice, starting from the initial local tensor $T^{(0)}$ \cite{XiePRB12,Exactblocking13prd,Yuxy13}, $2N$ HOTRG steps are applied in the two directions alternatingly to get a coarse-grained tensor corresponding to a system with volume $2^N\times 2^N$. At the $n$th step, firstly, a contracted tensor, $M^{(n)}$, is defined \cite{XiePRB12} by connecting two local tensors $T^{(n-1)}$:
\begin{equation}
\label{eq:mtensor}
M^{(n)}_{xx'yy'}=\sum_{i}T^{(n-1)}_{x_1x'_1yi}T^{(n-1)}_{x_2x'_2iy'},
\end{equation}
where $x=x_1\otimes x_2$ and $x'=x'_1\otimes x'_2$. Secondly, the new local tensor, $T^{(n)}$, is formed by applying an unitary transformation $U^{(n)}$ followed by a truncation to the two sides of $M^{(n)}$ with product states $x$ and $x'$,
\begin{equation}
\label{eq:ttensor}
T^{(n)}_{xx'yy'}=\sum_{ij}U^{(n)}_{ix}M^{(n)}_{ijyy'}{U^{*}}^{(n)}_{jx'}.
\end{equation} 
If $U$ is a real matrix, the parity symmetry is satisfied as
\begin{equation}
\label{eq:parity}
T_{xx'yy'}=T_{x'xyy'}.
\end{equation}

For each step, the unitary matrix $U$ is determined by taking the singular value decomposition of a specific matrix denoted as $Q$. By reserving the number of states to $D_s$, we mean keeping the eigenvectors corresponding to the first $D_s$ largest singular values of $Q$. The $T$ tensor is projected into $D_s$ new states in each direction without losing much information. In Ref \cite{XiePRB12}, for real $\beta$, $Q$ was chosen as
\begin{equation}
\label{eq:qmatrix}
Q\equiv M'{M'}^{\dagger}=U\Lambda U^{\dagger},
\end{equation}
where the matrix $M'_{x,x'yy'}$ is converted from the tensor $M_{xx'yy'}$ by regrouping its indices $x'yy'$ into a single one. 

For complex $\beta$, however, the parity symmetry Eq. (\ref{eq:parity}) is broken if $U$ has complex entries. To keep the symmetry, we need to use orthogonal transformations. The first candidate of $Q$ matrix is the positive-definite symmetric matrix:      
\begin{equation*}
\label{eq:q1matrix}
{\rm Re}(M'{M'}^{\dagger})={\rm Re}M'{{\rm Re}M'}^{{\rm T}}+{\rm Im}M'{{\rm Im}M'}^{{\rm T}},
\end{equation*}
as it has the same trace as $M'{M'}^{\dagger}$. However, the $Q$ matrix is not necessarily positive-definite as the singular values are $|\lambda|$ for a symmetric matrix and $|i\lambda|$ for an antisymmetric matrix, where $\lambda$ are the eigenvalues. Then, the $Q$ matrix can be taken as these three special cases:
\begin{eqnarray*}
\label{eq:q2matrix}
{\rm Re}(M'{M'}^{{\rm T}})={\rm Re}M'{{\rm Re}M'}^{{\rm T}}-{\rm Im}M'{{\rm Im}M'}^{{\rm T}},\\
{\rm Im}(M'{M'}^{\dagger})={\rm Im}M'{{\rm Re}M'}^{{\rm T}}-{\rm Re}M'{{\rm Im}M'}^{{\rm T}},\\
{\rm Im}(M'{M'}^{{\rm T}})={\rm Re}M'{{\rm Im}M'}^{{\rm T}}+{\rm Im}M'{{\rm Re}M'}^{{\rm T}}.
\end{eqnarray*}
At a fixed $D_s$, the distribution of the normalized singular values $\lambda_i/\lambda_1$  at different $\beta$ (e.g. Fig. \ref{fig:histograms}), which is associated with the error of calculation, are very similar to each other for these four cases except that for real $\beta$, the last two constructions are not present. The numerical results of all these four constructions are comparable with each other and better than the MC reweighting calculation. From the comparison between the results with the exact solution of the 2D Ising case, the errors from $Q={\rm Re}(M'{M'}^{{\rm T}})$ are smoother than the other three cases. For this reason, the calculations below were done with $Q={\rm Re}M'{M'}^{{\rm T}}$. We have checked in several cases that other choices lead to similar results.

\section{The Ising model with complex $\beta$}
\label{sec:ising}
We first consider the 2D Ising model on $L\times L$ lattice with even $L$,
\begin{equation}
\beta H=-\beta\sum_{<ij>}S_i\cdot S_j.
\end{equation}
The exact solution for the partition function of this model at finite volume and periodic boundary conditions was written by Kaufman in 1949 \cite{Kaufman1949is}. When $\beta$ is complex, the choices of signs for the square roots are discussed in the Appendix. For even $L$, the Hamiltonian is always a multiple of four and 
\begin{equation}
\label{eq:periodic1}
Z(\beta)=Z(\beta+i n\pi/2).
\end{equation}

\begin{figure}[h]
\advance\leftskip -1cm
 \includegraphics[width=4.3in]{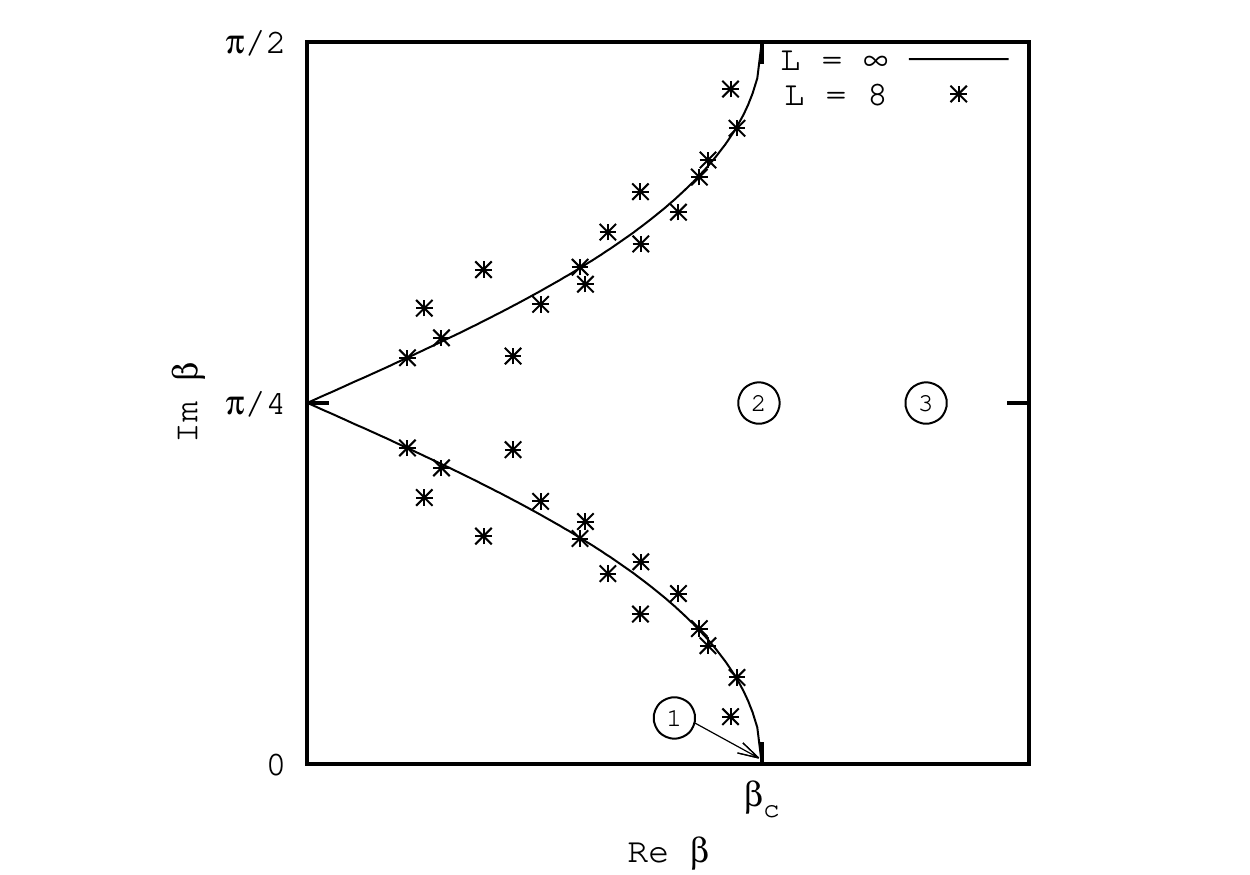}
 \caption{\label{fig:infvzero} Fisher curves and the zeros at finite volume for a $L=8$ system. Complex $\beta$ in region 1, 2, and 3 are displayed.}
 \end{figure}

In the infinite volume limit, the partition function zeros lie on two circles in the complex $\tanh\beta$ plane given by $\pm 1+\sqrt{2}\exp(i\theta)$ ($0\leq\theta\leq 2\pi$) \cite{fisher65}, mapping to two curves (we call them ``Fisher curves") each with periodicity $\pi$ in the imaginary direction of the complex $\beta$ plane. The first quadrant part of the zeros with $0\leq{\rm Im}\beta\leq \pi/2$ are shown in Fig. \ref{fig:infvzero}. The reason why the finite volume zeros are not exactly on the Fisher curves is explained at the end of the Appendix. The zeros in the other quadrants are just the mirror images along x-axis and y-axis given that $Z(-\beta)=Z(\beta)$ and $Z(\beta^*)=Z(\beta)^*$.

In this section, we use the HOTRG method to calculate the partition function $Z(\beta)$ and the free energy for the finite volume model. As the partition function is huge ($\sim e^{aL^2}$), a normalized version as in Eq. (\ref{eq:normpart}) needs to be used. For large ${\rm Im}\beta$, however, Eq. (\ref{eq:normpart}) may give results too small to analyze and a different normalizing factor $Z(\beta_0)$ is chosen, in which $\beta_0$ can be complex as long as $Z(\beta_0)$ is not zero. To analyze the accuracy of the TRG results, we calculate the relative error of the real part of free energy in logarithmic scale, namely, minus significant digits (-S.D.) compared to the exact solution, 
\begin{equation}
\label{eq:sd}
\text{S.D.}=-\log_{10}\left|\frac{f_{\text{HOTRG}}-f_{\text{exact}}}{f_{\text{exact}}}\right|.
\end{equation}
By looking at the comparison of the results from the HOTRG calculation with the exact solution for different $\beta$, we find that the accuracy of the HOTRG calculation is related to the distance from the $\beta$ to zeros of partition function. The closer to any zero, the larger $D_s$ is needed to get a more reliable result. Taking a $64\times 64$ lattice system for example, we consider $\beta$s on short line segments in three different regions (Fig. \ref{fig:infvzero}). Region 1: ${\rm Re}\beta=0.437643$, ${\rm Im}\beta\in[0.012,0.014]$, in which the approximated lowest zero $0.437643+i0.01312$ is included. Region 2: ${\rm Re}\beta=0.437643$, ${\rm Im}\beta\in[0.784,0.786]$, in which the point $0.437643+i\pi/4$ with the largest distance to zeros on the line ${\rm Re}\beta=0.437643$ is included. The HOTRG calculation for different periodical factor $n$ in Eq. (\ref{eq:periodic1}) also confirms the periodicity of the partition function. Therefore only the result from $\beta$ in the first period are needed. Region 3: ${\rm Re}\beta=0.6$ and the same imaginary part range as Region 2. We also compare the HOTRG results with those from MC reweighting (69 $\beta$ in $[0.2,0.625]$, 500000 configurations for each $\beta$) and find that the TRG results are better than MC in all three regions.

\begin{figure}[h]
\advance\leftskip -0.7cm
 \includegraphics[width=4.3in]{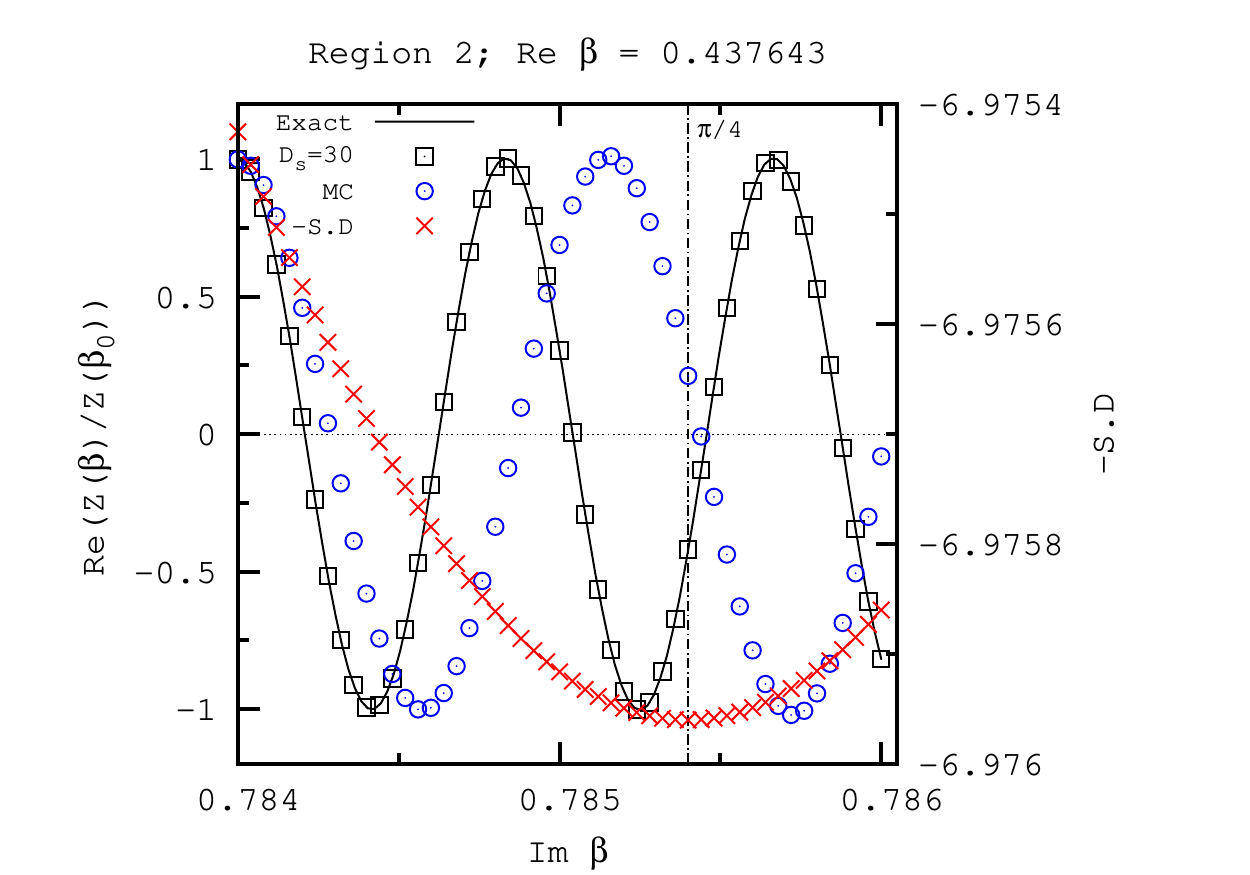}
 \caption{\label{fig:pr2}Left: The real part of the normalized partition function for region 2, result from the HOTRG with $D_s=30$, MC, and the exact solution. Right: relative error of the real part of free energy from HOTRG calculation, minimum at ${\rm Im}\beta=\pi/4$. }
 \end{figure}

\begin{figure}[h]
\advance\leftskip -0.7cm
 \includegraphics[width=4.3in]{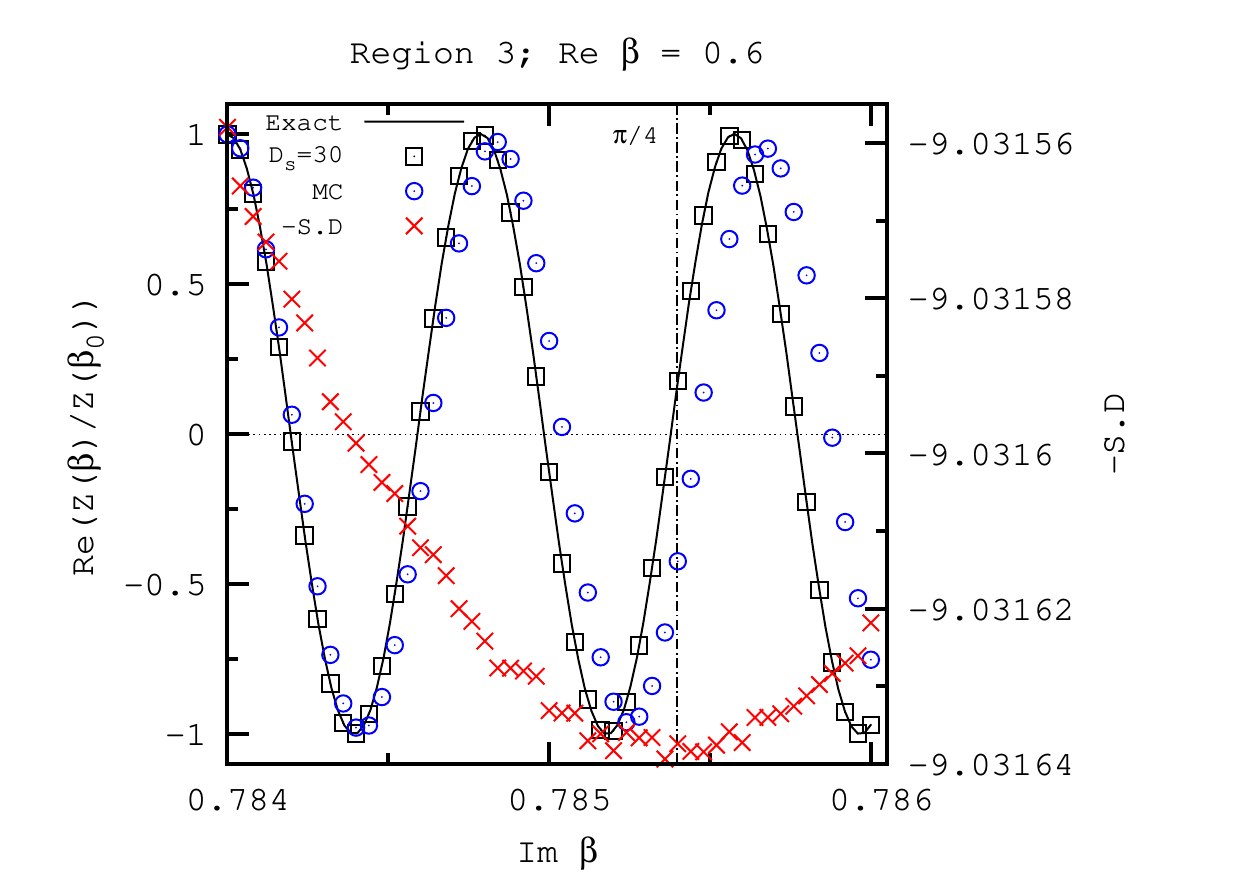}
 \caption{\label{fig:pr3} Left: The real part of the normalized partition function for region 3, result from HOTRG with $D_s=30$, MC, and the exact solution. Right: relative error of the real part of free energy from HOTRG calculation, minimum of the error curve at ${\rm Im}\beta=\pi/4$ approximately. }
 \end{figure}

As examples for the case where $\beta$ is far from any zero (region 2 and 3), Fig. \ref{fig:pr2} and \ref{fig:pr3} shows the real part of the normalized partition function $Z(\beta)/Z(\beta_0)$, in which $\beta_0=0.437643+i0.784$ for region 2 and $\beta_0=0.6+i0.784$ for region 3. Comparing the -S.D. in Fig. \ref{fig:pr2} and \ref{fig:pr3}, it is clear that the $\beta$s in region 3, farther from zeros, have smaller error. In one region, the approximate minimum of the error curve is at ${\rm Im}\beta=\pi/4$, which has the largest distance to any zero.   

\begin{figure}[h]
\advance\leftskip -1cm
 \includegraphics[width=4.3in]{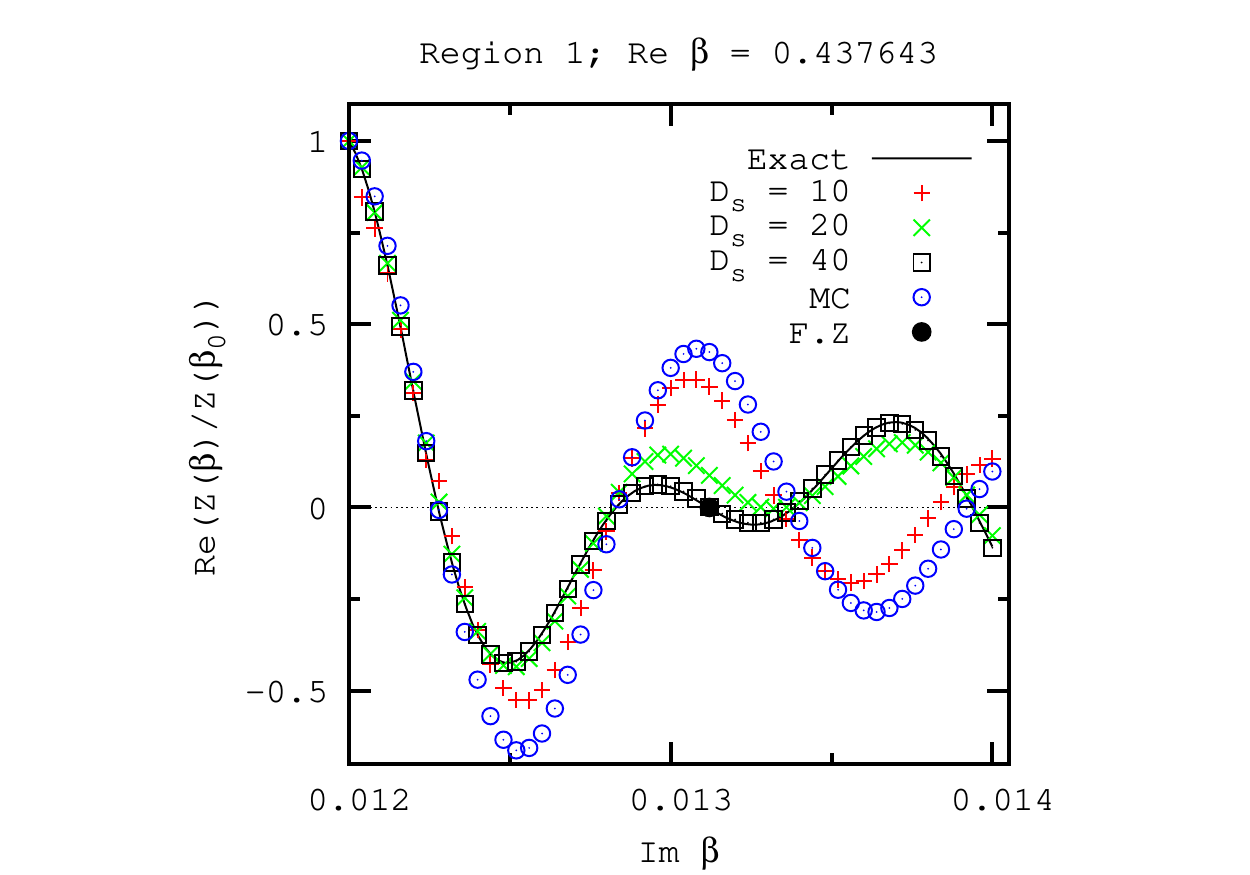}
 \caption{\label{fig:nearzero} The real part of the normalized partition function for $\beta$ near the Fisher zero $0.437643+i0.01312$ (the big filled circle on the real axis): result from the HOTRG with $D_s=10$, $20$, and $40$ ($D_s=30$ result is not shown as it is close to the $D_s=40$ case), MC, and exact solution.}
 \end{figure}

For region 1, where $\beta$ is in the vicinity of a Fisher zero, larger $D_s$ is needed to obtain results with a small error. Figure \ref{fig:nearzero} shows ${\rm Re}[Z(\beta)/Z(\beta_0)]$ ($\beta_0=0.437643+i0.012$) for $\beta$s in region 1. By increasing $D_s$, the HOTRG results get closer to the exact solution. With current configurations, the MC reweighting results are worse than the HOTRG results with $D_s=10$ near the lowest zero. 

\begin{figure}[h]
\advance\leftskip -1cm
 \includegraphics[width=4in]{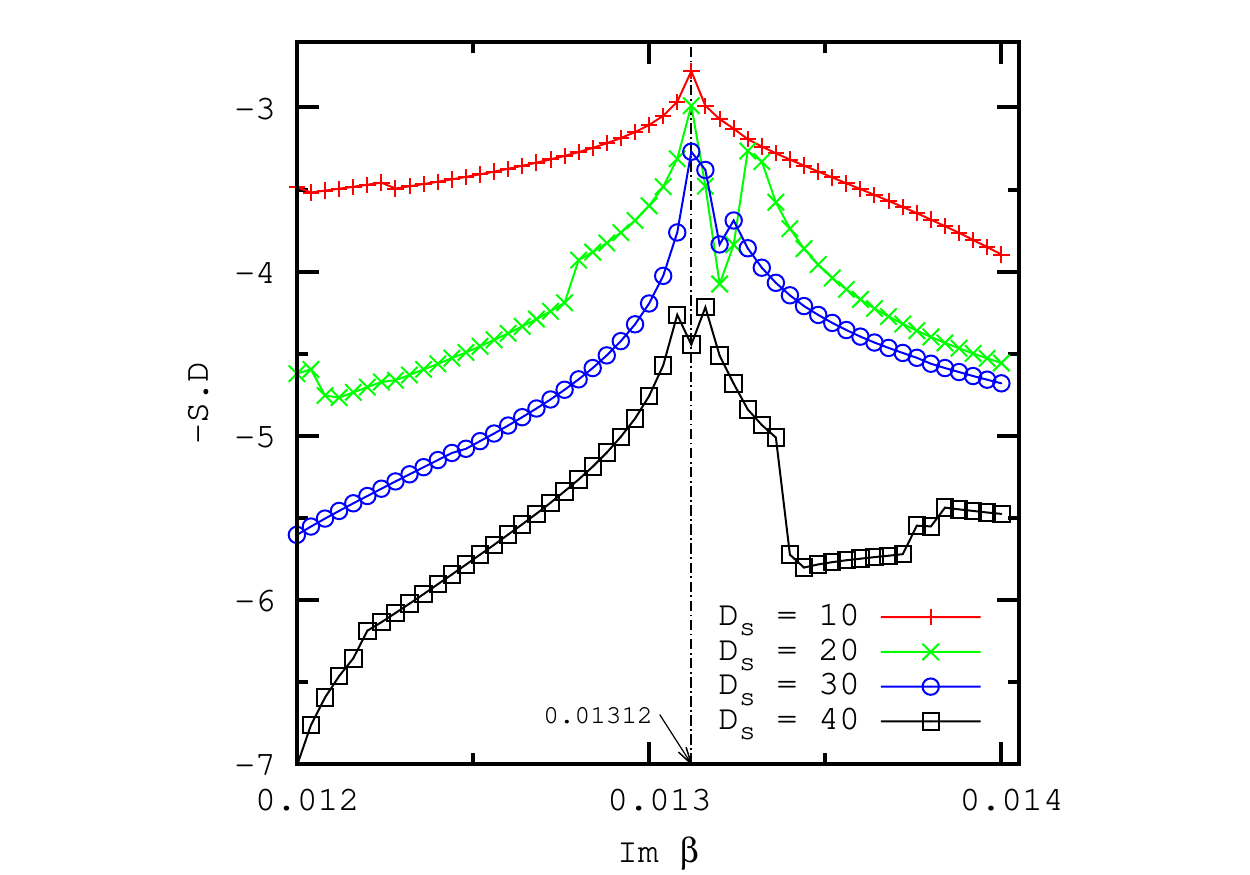}
 \caption{\label{fig:errorzero} The relative error of the real part of free energy for HOTRG calculation with $D_s=10$, 20, 30, and 40. Vertical line corresponds to the lowest zero.}
 \end{figure}

\begin{figure}[h]
\advance\leftskip -1cm
 \includegraphics[width=4.2in]{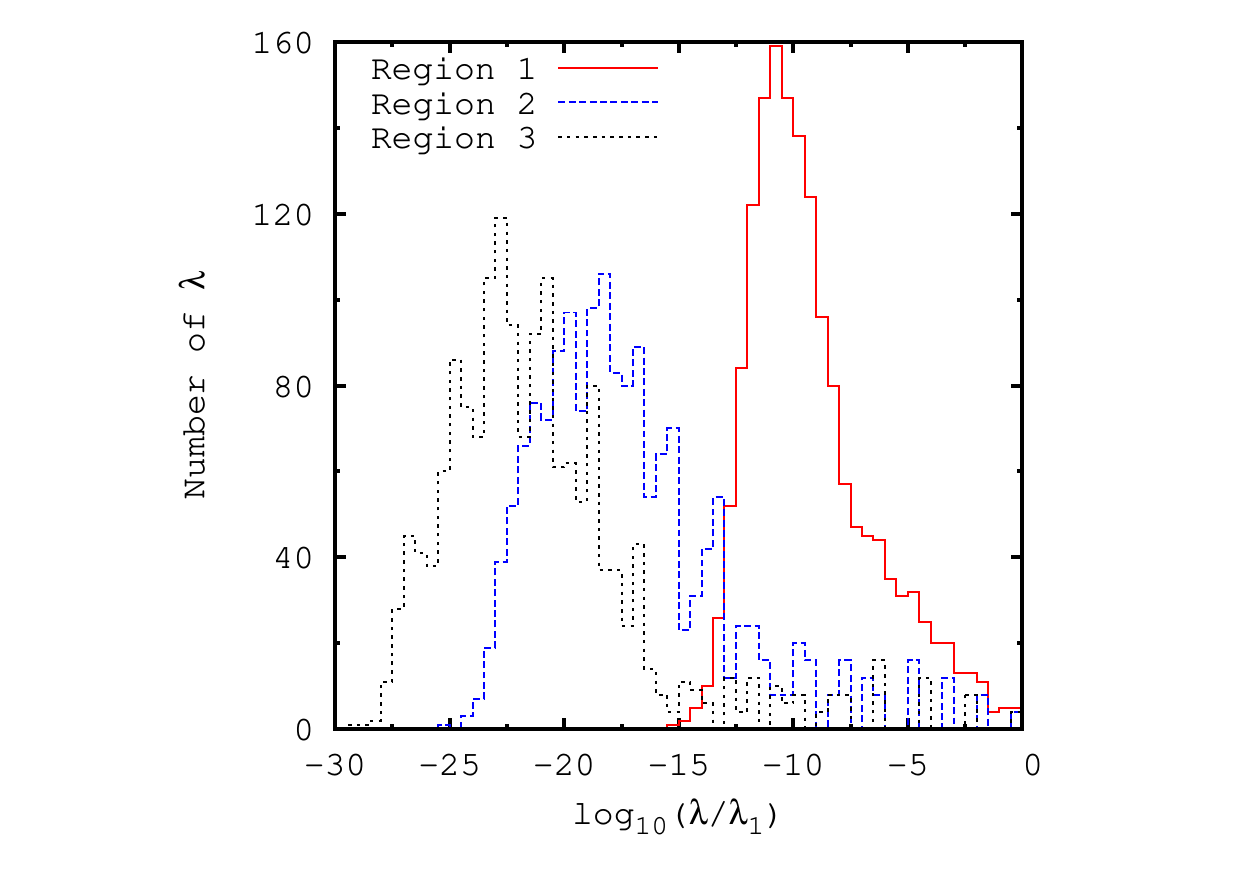}
 \caption{\label{fig:histograms} The distributions of normalized singular values $\lambda/\lambda_1$ for $\beta_0$ in region 1, 2, and 3, with $D_s=40$. There are 1600 singular values for each case.}
 \end{figure}

To understand the error of the HOTRG calculation near the lowest zero, we plot the relative error of the real part of free energy in Fig. \ref{fig:errorzero} for different $D_s$. At fixed $D_s$, the closer to the zero, the larger the error, and with fluctuation due to the fact that it is near the singularity point of the free energy at Fisher zeros.  
Besides the relative error of the free energy, the distributions of all $D^2_s$ normalized singular values of the $Q$ matrix are different for $\beta$s in the three regions at fixed $D_s$. In Fig. \ref{fig:histograms}, The peak of the histogram decreases (move to left) as the $\beta$ changes from region 1 to region 3. This shows that the truncated $D^2_s-D_s$ singular values are smaller when $\beta$ are farther away from the Fisher zeros.   

\begin{figure}[h]
 \begin{center}
 \includegraphics[width=2.85in]{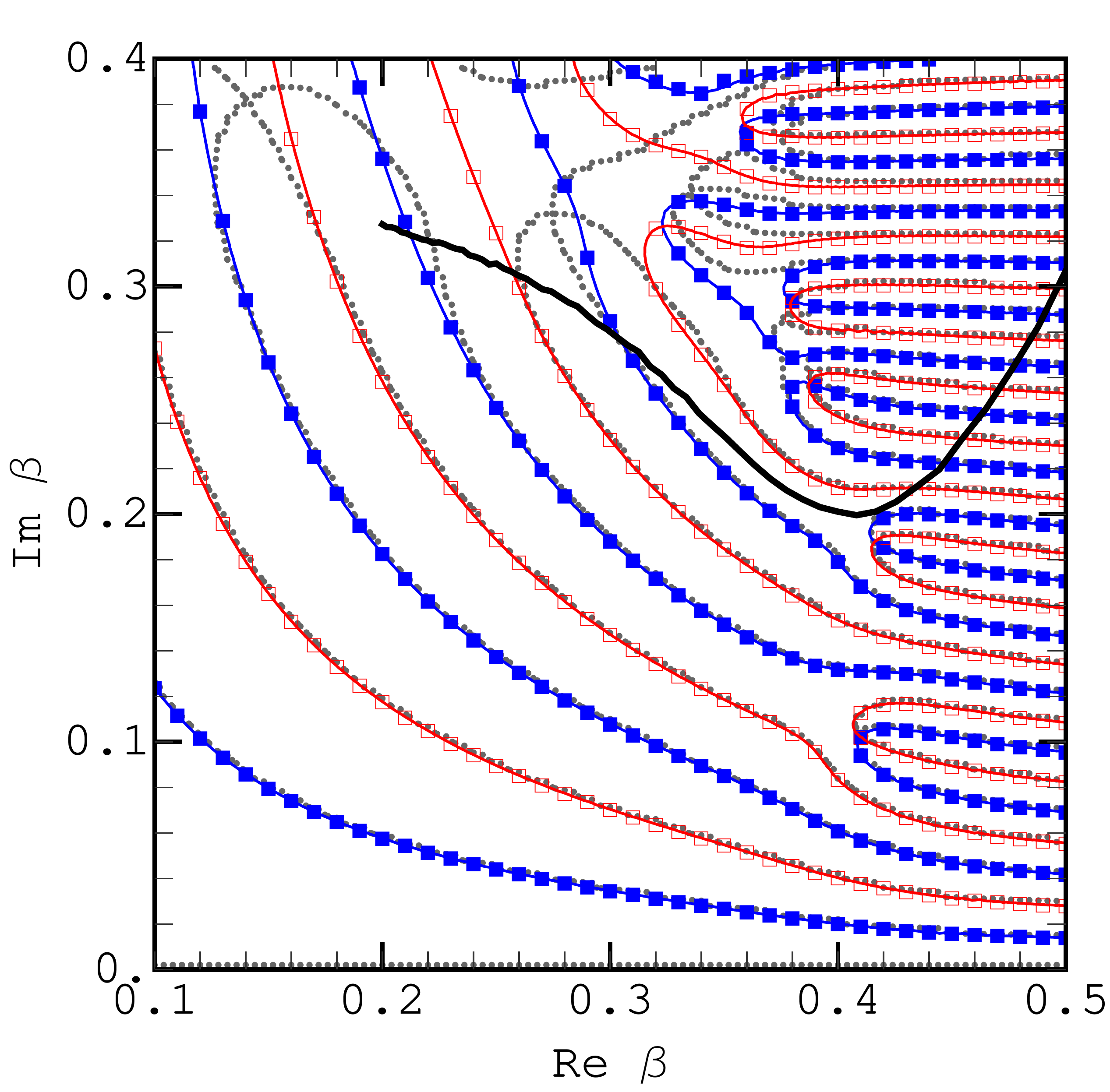}
 \caption{\label{fig:8x8zerosD40} Zeros of Real (\textcolor{blue}{$\blacksquare$}) and Imaginary (\textcolor{red}{$\Box$}) part of the partition function of Ising model at the volume $8\times 8$ from the HOTRG calculation with $D_s=40$ are on the exact solution lines. Gray dots: MC reweighting solution. Thick Black curve: the region of confidence for the MC reweighting result, above this line, the MC error is large.}
 \end{center}
 \end{figure}

The direct calculation of the partition function with complex $\beta$ provides a way for one to search for the zeros of the partition function. To locate the zeros, we scan the complex $\beta$ plane to find two sets of curves where the real and imaginary parts of the partition function are zeros. By using small sweep span, we can construct the contours corresponding to zero value for the real and imaginary parts. The intersection between the real and imaginary contours are the complex partition function zeros. In Fig. \ref{fig:8x8zerosD40}, the zeros for a $8\times 8$ system from the HOTRG with $D_s=40$ is shown, from which one can see that the HOTRG calculation reproduces the exact solution, while the MC calculation has larger errors at large imaginary $\beta$. To estimate the MC results, the region of confidence $\Delta\beta$ is calculated to locate the largest reliable ${\rm Im}\beta$ where the fluctuation has the same size of the average \cite{PhysRevD.76.116002}.
\begin{equation}
\label{eq:roc}
|\Delta\beta|^2<\ln(N_{\rm conf.}/\tau)/\sigma^2,
\end{equation}
where $N_{\rm conf.}$ is the number of configurations, $\tau$ is the integrated correlation time, and $\sigma$ is the standard deviation of the total action and scales like $V^{1/2}$. $\Delta\beta$ is inversely proportional to the linear size of the system ${\rm Im}\beta_\text{max}\sim L^{-1}$ for 2D Ising model as $\nu=1$.
 
\section{The $O(2)$ model with complex $\beta$}
\label{sec:xy}
In this section, we will apply the HOTRG method to the 2D $O(2)$ model (the XY model) with complex $\beta$ on $L\times L$ lattice with even $L$, 
\begin{equation}
\beta H=-\beta\sum_{<ij>}\cos(\theta_i-\theta_j).
\end{equation}

\begin{figure}[h]
\advance\leftskip -1cm
 \includegraphics[width=4in]{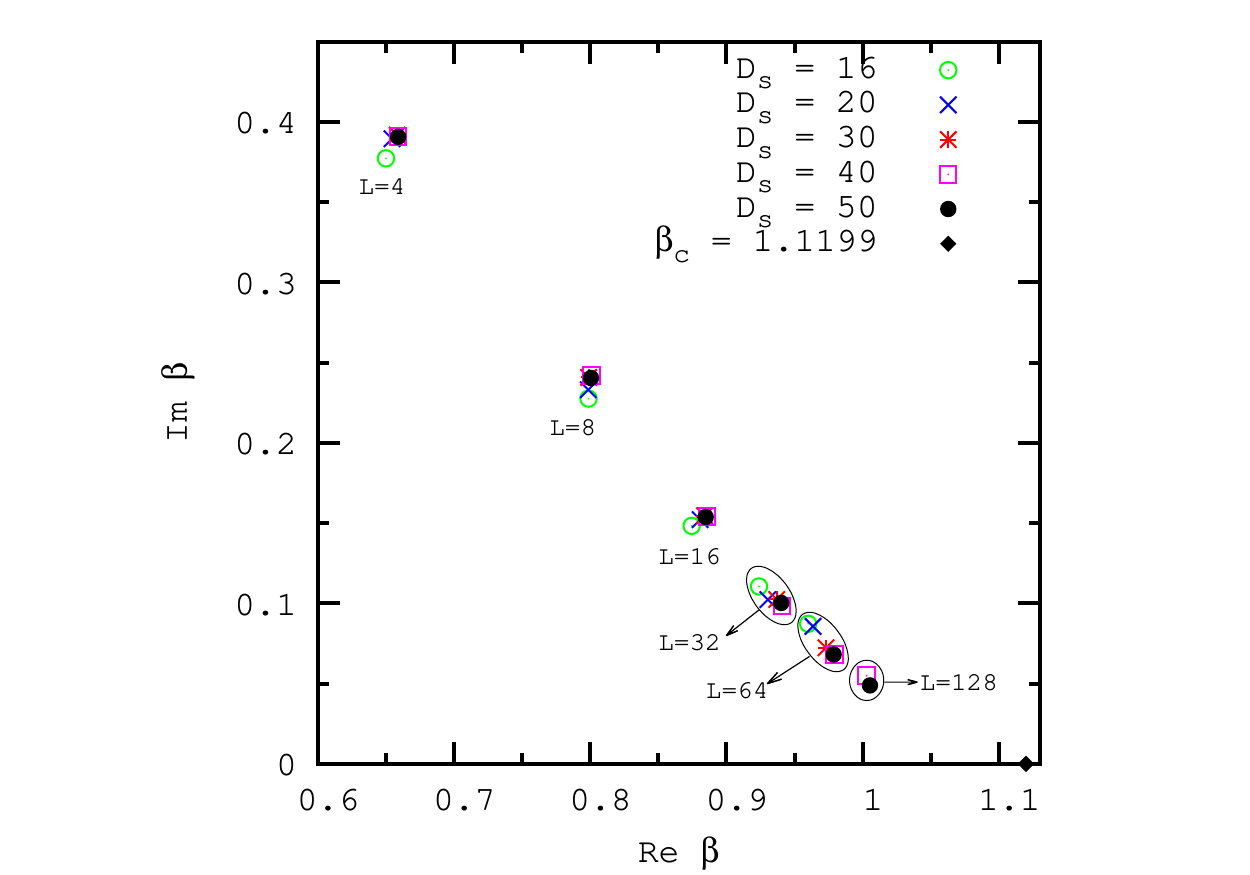}
 \caption{\label{fig:xyzeros} Fishers zeros of XY model with length $L=4$, 8, 16, 32, 64, and 128 (from up-left to down-right) at different $D_s$. For $L=128$, only $D_s=40$ and $50$ are shown.}
 \end{figure}

Following the search method introduced in the previous section, we can also locate the zeros of the 2D $O(2)$ model in finite volume. In Fig. \ref{fig:xyzeros}, the lowest zeros (listed in Table \ref{tb:zedsmall}, \ref{tb:zed450}) for different volumes from the HOTRG with different $D_s$ are shown. For each $L$, as $D_s$ increases up to 50, the zeros converge to a point in the complex $\beta$ plane with an error of order 0.01 for both the real and imaginary parts. 

\begin{figure}[h]
\advance\leftskip -1cm
 \includegraphics[width=4in]{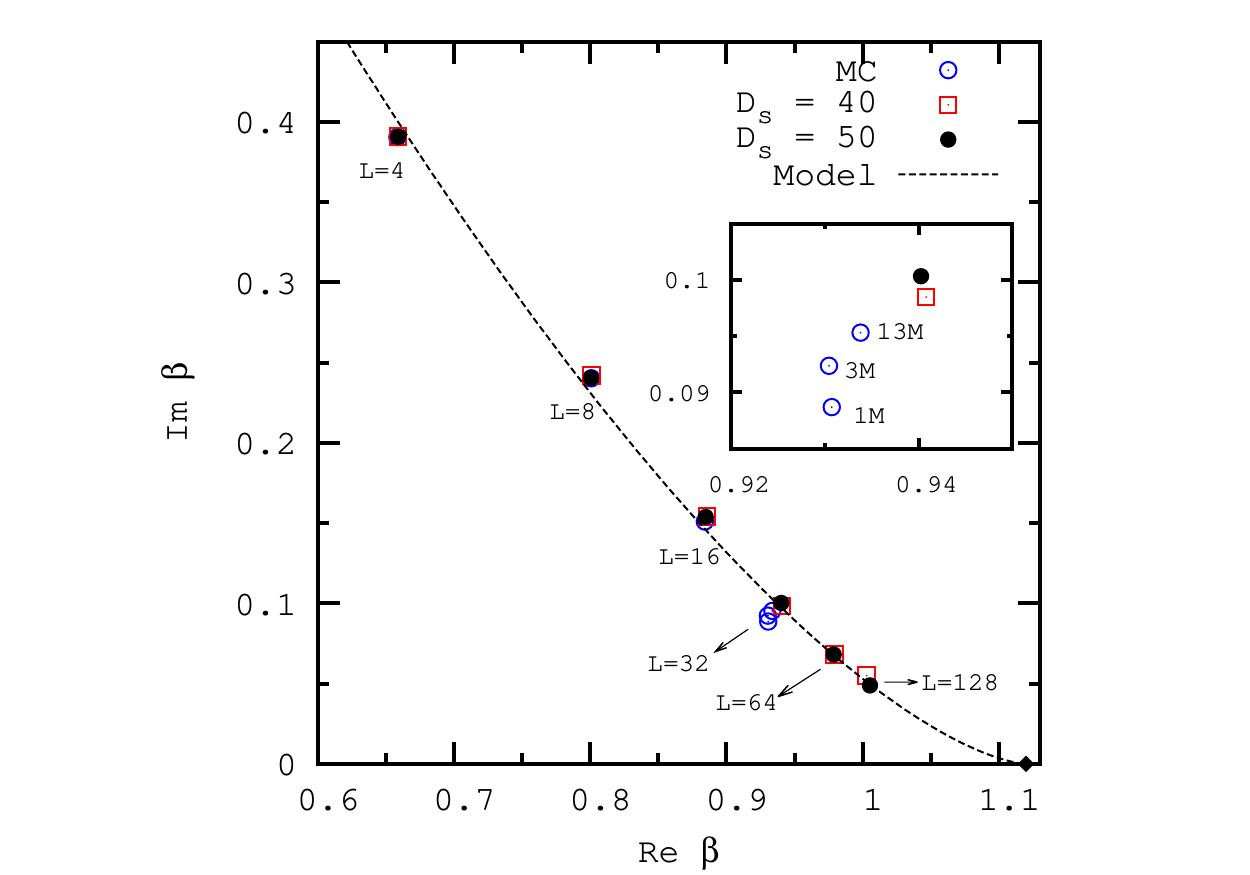}
 \caption{\label{fig:xyzwmc} Zeros of XY model with linear size $L=4$, 8, 16, 32, 64, and 128 (from up-left to down-right) calculated from HOTRG with $D_s=40$, and 50 and zeros with $L=4$, 8, 16, and 32 from MC. The curve is a model for trajectory of the lowest zeros.}
 \end{figure}

By using small sweep span in the complex $\beta$ plane, we locate the zeros with more digits for the cases of $D_s=40$ and $50$, results are shown in Table \ref{tb:zed450} (except for $L=128, D=40$, where the fluctuations are larger). Figure \ref{fig:xyzwmc} shows the comparison of the values from the HOTRG with $D_s=40$ and $50$ and zeros calculated from MC reweighting (from 41 $\beta$ in the region $[0.7,1.0]$). At small volume ($L=4$, 8, and 16), two results agree with each other within one percent. However at $L=32$, when the number of MC configurations for each $\beta$ was increased from $10^6$ to $1.3\times 10^7$, the MC results converged slower than the HOTRG calculations from $D_s=40$ to $50$.   
\begin{table}
\begin{center}
\begin{tabular}{|c|c|c|c|}
\hline
\hline
$L$&$D_s=16$&$D_s=20$&$D_s=30$\cr
\hline
4&$0.6501+i0.3774$&$0.6545+i0.3896$&$0.6585+i0.3919$\cr
\hline
8&$0.7988+i0.2277$&$0.7985+i0.2332$&$0.7989+i0.2409$\cr
\hline
16&$0.8745+i0.1483$&$0.8807+i0.1522$&$0.8841+i0.1547$\cr
\hline
32&$0.9239+i0.1106$&$0.9304+i0.1024$&$0.9370+i0.1025$\cr
\hline
64&$0.9600+i0.08728$&$0.9636+i0.08565$&$0.9731+i0.07235$\cr
\hline
\hline
\end{tabular}
\end{center}
\caption{\label{tb:zedsmall} The lowest zeros from HOTRG calculation with $D_s=16$, 20, and 30 for different volumes.}
\end{table}

\begin{table}
\begin{center}
\begin{tabular}{|c|c|c|}
\hline
\hline
$L$&$D_s=40$&$D_s=50$\cr
\hline
4&$0.658866+i0.3909946$&$0.6590215+i0.3907913$\cr
\hline
8&$0.801055+i0.2419952$&$0.8006218+i0.2405708$\cr
\hline
16&$0.885983+i0.1542734$&$0.8848648+i0.1538418$\cr
\hline
32&$0.940816+i0.09848932$&$0.9402603+i0.10032925$\cr
\hline
64&$0.979298+i0.0680147$&$0.9787804+i0.06812091$\cr
\hline
128&$1.003+i0.05492$(noise)&$1.0054329+i0.0488771$\cr
\hline
\hline
\end{tabular}
\end{center}
\caption{\label{tb:zed450} The lowest zeros from the HOTRG calculation with $D_s=40$ and $50$ for different volumes (except for $D_s=40$, $L=128$).}
\end{table}

We performed finite size scaling for the lowest zeros calculated from the HOTRG with $D_s=50$ (Table \ref{tb:zed450}) to estimate the behavior of the lowest zeros for larger $L$. For one step of the RG transformation with scaling factor $\tilde{b}$ at large $L$ ($L\rightarrow L/\tilde{b}$), the correlation lengh scales like $\xi\rightarrow \xi/\tilde{b}$. By assuming that the singular part of the partition function is a function $f(\xi/L)$, then at the zeros,   
\begin{equation}
Z(\beta_{\text{z}})=f(z_0)=0,
\end{equation}
the values of zeros for different volumes map to the same $z_0$ at large $L$. 

From the scaling behavior of the correlation length for a Kosterlitz-Thouless (K-T) transition \cite{KTXY1973,Kosterlitz1974},
\begin{equation}
\xi=A\exp(b/\sqrt{t}),
\end{equation}
where $t=\beta_{\text{c}}-{\rm Re}\beta-i{\rm Im}\beta$ for ${\rm Re}\beta < \beta_{\text{c}}$ and $\beta_c$ is the K-T transition coupling. By taking the imaginary part as a perturbation from the real part and from the leading order equation for real and imaginary $t$, one can obtain the relation between ${\rm Re}\beta_{\text{z}}$ and ${\rm Im}\beta_{\text{z}}$ as
\begin{equation}
\label{eq:zmodel}
{\rm Im}\beta_{\text{z}}=\frac{2a}{b}(\beta_{\text{c}}-{\rm Re}\beta_{\text{z}})^{3/2},
\end{equation}
where $a={\rm Im}(\ln(z_0/A))$. 
  
According to the high accurate results for critical $\beta_c$ from Monte Carlo \cite{Hasenbusch:2005xm,Komura12gpu} and High Temperature expansion \cite{Butera:2008zz,Arisue09hte}, we fix the critical coupling $\beta_c=1.1199$ and do a one-parameter fit for the zeros $\beta_c$ from the HOTRG calculation with $D_s=50$. The best fitting parameter value is $a/b=0.63942\pm 0.00919$. From Fig. \ref{fig:xyzwmc}, the current zeros are very close to the model. It is obvious that the zeros pinch the real axis convexly as $L$ increases for the K-T transition, while in the case of the 2D Ising model which has a second-order phase transition the zeros pinch the real axis concavely.

We can also do more-parameter fits for the scaling relation of the real part of $t$ separately to obtain the critical coupling $\beta$, up to the second digits of which is consistent with the values obtained in these refs \cite{Hasenbusch:2005xm,Komura12gpu,Butera:2008zz,Arisue09hte}. However, to get more accurate critical coupling, zeros at larger volume with more $D_s$ are needed for the HOTRG calculation. The computational demands for time and memory of such calculations seem to require more than a laptop.   
 
\section{Conclusions}
\label{sec:conl}
In conclusions, we have shown that the partition function of spin models (2D Ising model and 2D $O(2)$ model) at complex coupling $\beta$ can be calculated accurately by the HOTRG method even at large ${\rm Im}\beta$ where the MC reweighting method fails. Reliable zeros of the partition function of these spin models can also be obtained by the HOTRG calculation. By the finite size scaling for the lowest zeros of 2D $O(2)$ models at different volumes, we have shown that the zeros will pinch the real axis convexly in the infinite volume limit for the K-T transition. 

In summary, the success of the HOTRG method to solve the complex partition function and obtain the zeros of the partition functions shows that this new method seems insensitive to the problem associated with complex value. This allows us to apply the TRG method to study another sign problem, the $Z(N)$ and $O(2)$ models with a complex chemical potential $i\mu$. Take the 2D $O(2)$ model 
\begin{equation}
\label{eq:chemicalp}
\beta H=-\beta\sum_{<ij>}\cos(\theta_i-\theta_j-i\mu)
\end{equation} 
as an example, the local tensor can be formulated by the modified Bessel function of the first kind ($I_n\equiv I_n(\beta)$) and $\exp(n\mu)$:
\begin{equation}
\label{eq:tchem}
T_{xx'yy'}=\sqrt{I_x e^{x\mu}}\sqrt{I_y e^{y\mu}}\sqrt{I_{x'} e^{x'\mu}}\sqrt{I_{y'} e^{y'\mu}}\delta_{x+y,x'+y'},
\end{equation}
which have no complex terms at all. We plan to compare the results from TRG calculation with those obtained with other method such as dual formulations \cite{Meisinger:2013zfa} and world-line methods \cite{Shailesh2010,worm2001} in the future.

\begin{acknowledgments}
This research was supported in part by the Department of Energy under Award Numbers DE-SC0010114 and FG02-91ER40664. We have used resources of the National Energy Research Scientific Computing Center which is supported by the Office of Science of the U.S. Department of Energy under Contract No. DE-AC02-05CH11231 and The University of Iowa's Helium Cluster. Y. L. was supported by the URA Visiting Scholars' program. Fermilab is operated by Fermi Research Alliance, LLC, under Contract No.~DE-AC02-07CH11359 with the United States Department of Energy. T. X., Z. X, J. Y, and M. Q were supported by the National Natural Science Foundation of China (Grants No.~10934008 and No.~10874215) and MOST 973 Project (Grant No.~2011CB309703). Our work on the subject started while attending the KITPC workshop ``Critical Properties of Lattice Models" in summer 2012. Y. M. did part of the work while at the workshop ``LGT in the LHC Era" in summer 2013 at the Aspen Center for Physics supported by NSF grant No 1066293. 
\end{acknowledgments}

\appendix*
\section{Kaufman's Exact Solution}
The partition function $Z(L,\beta)$ of a finite $L \times L$ square lattice with $L$ even and periodic boundary conditions is

\begin{equation}
Z(L,\beta)) = \frac{1}{2} (2 \sinh (2 \beta))^{\frac{L^2}{2}} \sum _{i=1}^4 Z_i(\beta),
\label{eq:partition}
\end{equation}
with 

\begin{align}
Z_1	&=	\prod _{r=0}^{L-1} 2 \cosh \left(\frac{1}{2} L \gamma_{2
   r+1}\right),\\
Z_2	&=	\prod _{r=0}^{L-1} 2 \sinh \left(\frac{1}{2} L \gamma_{2
   r+1}\right),\\
Z_3	&=	\prod _{r=0}^{L-1} 2 \cosh \left(\frac{1}{2} L \gamma_{2 r}\right),\\
Z_4	&=	\prod _{r=0}^{L-1} 2 \sinh \left(\frac{1}{2} L \gamma_{2 r}\right),
\label{eq:fourF}
\end{align}
where 

\begin{equation}
\cosh\gamma_l	=	c_l	=	\cosh (2 \beta) \coth (2 \beta)-\cos \left(\frac{\pi  l}{L}\right),
\label{eq:cosh}
\end{equation}
so that 

\begin{align}
\gamma_0	&=	2 \beta+\log (\tanh (\beta)),\\
\gamma_l	&=	\log \left(\sqrt{c_l^2-1}+c_l\right) , l \neq 0.
\end{align}
These expressions contain square roots and logarithms that can lead to cuts and discontinuities if not handled properly. We know that at finite volume, the partition function is an analytical function in the entire complex $\beta$ plane. 
More precisely, it is a sum of exponentials with integer weights that count the number of ways we can have a given energy. For instance, for $L=4$, the partition function reads:

\vspace{2mm}
$2 e^{-32 \beta }+32 e^{-24 \beta }+64 e^{-20 \beta }+424 e^{-16 \beta }+1728
   e^{-12 \beta }+6688 e^{-8 \beta }+13568 e^{-4 \beta }+13568 e^{4 \beta }+6688
   e^{8 \beta }+1728 e^{12 \beta }+424 e^{16 \beta }+64 e^{20 \beta }+32 e^{24
   \beta }+2 e^{32 \beta }+20524$.

\vspace{2mm}   
As we assumed $L$ even, the use of square roots and logs can be circumvented by expressing the factors in Eq. (\ref{eq:fourF}) in terms of the Chebychev polynomials:

\begin{eqnarray}
\cosh((L/2) \gamma _{r})&=& T_{L/2}(\cosh(\gamma_{r})), \\
\sinh((L/2) \gamma _{r})&=& U_{L/2-1}(\cosh(\gamma_{r})) \sinh(\gamma_{r}).
\end{eqnarray}
Using Eq. (\ref{eq:cosh}) and using the pre factor to cancel poles at $\beta = 0$, we see that the first and third terms of the partition function are now polynomials of entire functions. The factors $\sinh(\gamma_{r})$ 
have a sign ambiguity however they come in pairs except for $\sinh(\gamma_{0})$  and $\sinh(\gamma_{L})$. A careful reading of footnotes in Ref. \cite{Kaufman1949is} yields
\begin{eqnarray}
\sinh(\gamma_{0})&=&\cosh(2\beta)-\coth(2\beta),\\
\sinh(\gamma_{L})&=&\cosh(2\beta)+\coth(2\beta).
\end{eqnarray}
Combining these results, the partition function is clearly an entire function. We have checked that this procedure reproduces the exact results for the partition function and the zeros for different $L$. For example, from the location of zeros of a $8\times 8$ system in Fig. \ref{fig:infvzero}, one can find that the zeros of a finite volume system are still in the vicinity of the zero curves from infinite volume limit. It should be noted that locus of the zeros given by Fisher \cite{fisher65}, $\pm 1+\sqrt{2}\exp(i\theta)$ ($0\leq\theta\leq 2\pi$) represent curves where the zeros of the individual terms $Z_1$, $Z_2$, $Z_3$, and $Z_4$. However at finite volume, the zeros of the four terms are slightly different and the zeros of the sum of the four terms are slightly away from the Fisher curves.

\bibliography{sign.bbl}

\end{document}